\documentclass{aa}
\usepackage{amssymb,amsmath,amsfonts}
\usepackage{graphicx}
\usepackage{color}
\usepackage{ulem}             
\usepackage{multicol}
\usepackage{caption}
\usepackage{pgf,tikz}
\usetikzlibrary{arrows}
\usepackage{colortbl}

\usepackage{xcolor}

\usepackage{amssymb}
\usepackage{latexsym}
\usepackage{pstricks}
\usepackage{pst-plot}
\usepackage{tabularx}
\usepackage{mathtools}
\usepackage{bm}
\usepackage{soul}
\usepackage{cancel}

	\usetikzlibrary[patterns]

\newcommand{\be}{\begin{equation}}
\newcommand{\ee}{\end{equation}}

\newcommand*\samethanks[1][\value{footnote}]{\footnotemark[#1]}

 {\everymath{\displaystyle\everymath{}}\array}%
 {\endarray}

\begin{document}

\title{Co-orbital exoplanets from close period candidates: The TOI-178 case
}
\subtitle{ }
\titlerunning{Coorbital candidates from close period pairs of exoplanets }

\author{A. Leleu$^1$\thanks{\email{adrien.leleu@space.unibe.ch}}\thanks{CHEOPS fellows}, J. Lillo-Box$^2$, M. Sestovic$^3$, P. Robutel$^4$, A. C. M. Correia$^{5,4}$, N. Hara$^6$\samethanks, D. Angerhausen$^{3,7}$, S. L. Grimm$^3$,  J. Schneider$^8$
}
\authorrunning{A. Leleu, et al.}

\institute{
$^1$ Physikalisches Institut, Universit\"at Bern, Gesellschaftsstr.\ 6, 3012 Bern, Switzerland.\\
$^2$ European Southern Observatory, Alonso de Cordova 3107, Vitacura Casilla 19001, Santiago 19, Chile. \\
$^3$ Center for Space and Habitability, University of Bern, Gesellschaftsstr.\ 6, 3012 Bern, Switzerland. \\
$^4$ IMCCE, Observatoire de Paris - PSL Research University, UPMC Univ. Paris 06, Univ. Lille 1, CNRS,
77 Avenue Denfert-Rochereau, 75014 Paris, France.\\
$^5$  CFisUC, Department of Physics, University of Coimbra, 3004-516 Coimbra, Portugal.\\
$^6$ Observatoire de Gen\`eve, Universit\'e de Gen\`eve, 51 ch. des Maillettes, 1290 Versoix, Switzerland.\\
$^7$ Blue Marble Space Institute of Science, 1001 4th Ave Suite 3201, Seattle, WA 98154, USA.\\
$^8$ Paris Observatory, LUTh UMR 8102, 92190 Meudon,France.\\
}

\abstract
{
Despite the existence of co-orbital bodies in the solar system, and the prediction of the formation of co-orbital planets by planetary system formation models, no co-orbital exoplanets (also called trojans) have been detected thus far. Here we study the signature of co-orbital exoplanets in transit surveys when two planet candidates in the system orbit the star with similar periods. Such pair of candidates could be discarded as false positives because they are not Hill-stable. However, horseshoe or long libration period tadpole co-orbital configurations can explain such period similarity. This degeneracy can be solved by considering the Transit Timing Variations (TTVs) of each planet.
We then focus on the three planet candidates system TOI-178: the two outer candidates of that system have similar orbital period and had an angular separation near $\pi/3$ during the TESS observation of sector 2. 
%
Based on the announced orbits, the long-term stability of the system requires the two close-period planets to be co-orbitals.
Our independent detrending and transit search recover and slightly favour the three orbits close to a 3:2:2 resonant chain found by the TESS pipeline, although we cannot exclude an alias that would put the system close to a 4:3:2 configuration.
We then analyse in more detail the co-orbital scenario. We show that despite the influence of an inner planet just outside the 2:3 mean-motion resonance, this potential co-orbital system can be stable on the Giga-year time-scale for a variety of planetary masses, either on a trojan or a horseshoe orbit. We predict that large TTVs should arise in such configuration with a period of several hundred days. We then show how the mass of each planet can be retrieved from these TTVs.
}


\maketitle

\section{Introduction}

Among the known multiplanetary systems, a significant number contain bodies in (or close to) first and second order mean-motion resonances (MMR) \citep{Fabrycky2014}. However, thus far no planets were found in a $0^{\rm th}$ order MMR, also called trojan or co-orbital configuration, despite several dedicated studies \citep{MaWi2009,Ja2013,HiAn2015b}, and the TROY project \citep{LiBo2017,LiBo2018}. 

Such bodies are numerous in the solar system, such as Jupiter and Neptune's trojans, or some of the Saturnian satellites. Although most of the time the mass of one of the co-orbitals is negligible with respect to the other, Janus and Epimetheus (co-orbital moons of Saturn) only have a mass ratio of 3.6. Indeed, trojan exoplanets
can be stable on durations comparable to the lifetime of a star as long as $(m_1+m_2)<m_0/27$, where $m_1$ and $m_2$ are the mass of the co-orbitals and $m_0$ the mass of the star \citep{Ga1843}. That implies that even the two most massive planets of our solar system, Jupiter and Saturn, could share the same orbital period with difference of mean longitudes librating around $60^\circ$. The less massive the two co-orbitals are, the larger their amplitude of libration around the $L_4/L_5$ equilibria can be. When both masses are similar to the mass of Saturn or lower, the two co-orbitals can also be on a stable horseshoe orbit, in which the difference of the mean longitudes librates with an amplitude of more than $310^\circ$ \citep{Gar1976b,Ed1977,NiPoRo2018}. Eccentric/inclined orbits offer a wealth of other stable configurations that are extensively studied \citep{NaMu2000,GiuBeMiFe2010,MoNa2013,RoPo2013,LeRoCo2018}. Trojan exoplanets are a by-product of our understanding of planetary system formation \citep{CreNe2008} and can form through planet-planet scattering or \textit{in-situ} accretion at the Lagrangian point of an existing planet \citep{LauCha2002}. {However, there are currently only few constrains on the expected characteristics (such as the amplitude of libration) of co-orbital exoplanets, due to the complexity of the evolution of such configuration in a protoplanetary disc \citep{CreNe2009,GiuBe2012,PiRa2014,LeCoAt2019}.}

The detection of co-orbital exoplanets is challenging due to the existence of degeneracies with other configurations across various detection techniques, Transit Timing Variations (TTVs) \citep{Ja2013,VoNe2014}, radial velocities and astrometry \citep{LauCha2002,GiuBe2012,LeRoCo2015}; while the transit of both planets require close-in coplanar systems. The multi-planet systems Kepler-132, Kepler-271 and Kepler-730 were first announced to contain close period planets until a more detailed analysis disfavoured the co-orbital scenario in favour of the planets orbiting two different stars of a binary (Kepler-132), or in a 2:1 mean-motion resonance \citep[Kepler-271 and Kepler-730,][]{Lissauer2011,Lissauer2014}.

 In this study we focus on (quasi-)coplanar orbits, as we aim to describe potential signals in the data from past and current transit surveys such as Kepler/K2 and TESS \citep{KEPLER,TESS}  and prepare for the analysis of future missions such as CHEOPS and PLATO \citep{CHEOPS,PLATO,HiAn2015}. After a brief summary on the dynamics and stability of close-period planets, in section \ref{sec:general} we discuss how TTVs can be used to remove the degeneracy between co-orbitals and seemingly similar but distinct orbital periods when candidates do not have a good enough signal to noise ratio to identify each transit individually.
In section \ref{sec:TOI} we consider the case of the TOI-178 system, where two of the announced three planet candidates have close orbital periods.
We first perform an independent detrending of the lightcurve and search for transits. Then, assuming that the estimated periods from the TESS data validation report are correct, we perform a stability analysis of this 3-planet system as a function of the masses of the planets, and predict the TTVs that should be observed in such system.


\section{Coorbital dynamics and stability}
\label{sec:general}
\subsection{Coorbital motion}
\label{sec:dyn}
 \begin{figure}
\begin{center}
\includegraphics[width=1\linewidth]{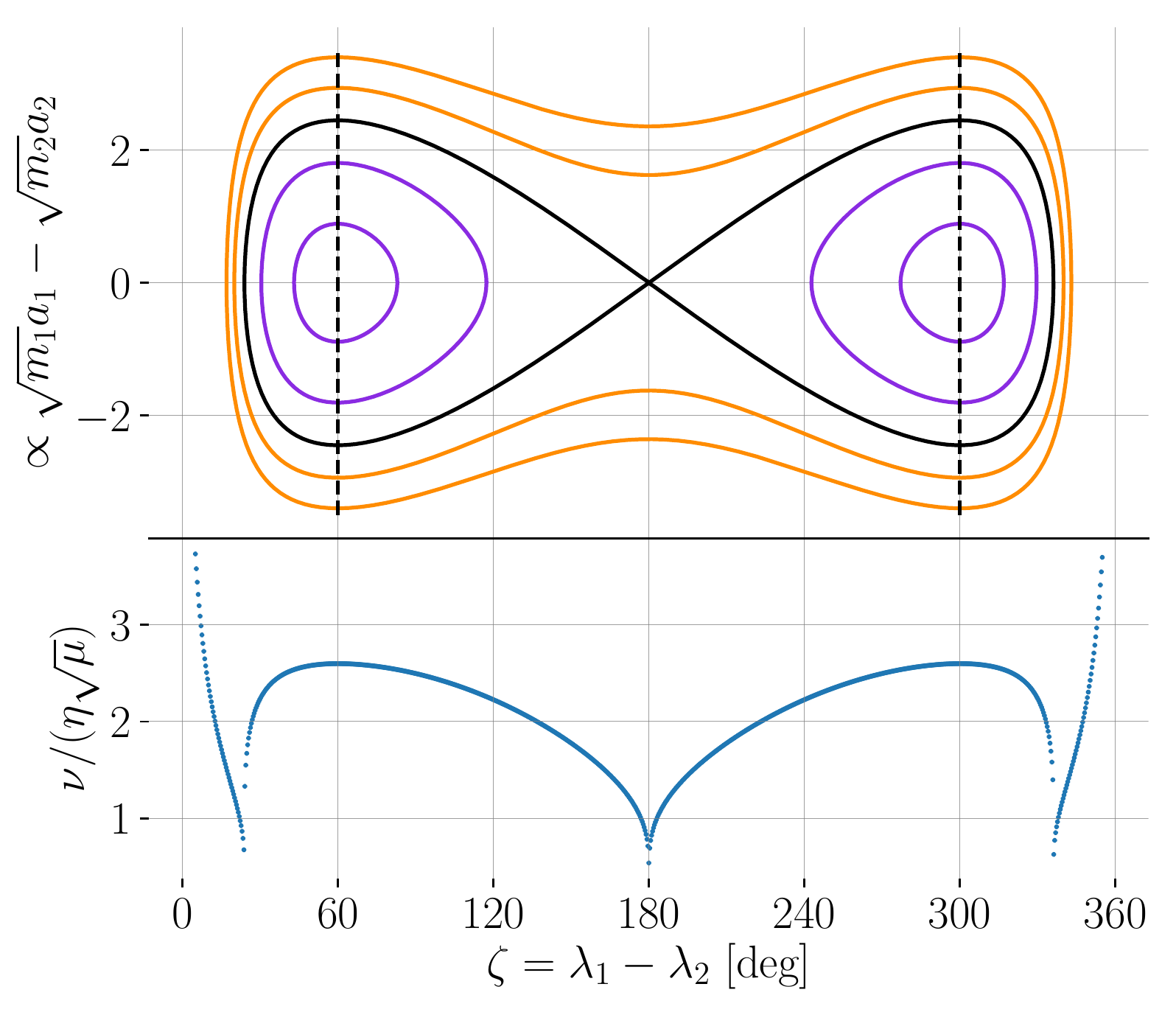}\\
\caption{\label{fig:zeta} Top: phase space of the co-orbital resonance valid for small eccentricities and inclinations. The x axis displays the resonant angle $\zeta=\lambda_1-\lambda_2$ while the y axis is its normalised angular frequency, which is proportional to $m_1\sqrt{a_1}-m_2\sqrt{a_2}$. Bottom: value of the normalised fundamental frequency $\nu$ for initial conditions along the line ($\zeta$,$\dot \zeta=0$) of the top graph.  Both graphs were obtained by integrating eq. (\ref{eq:eqerdi}).}
\end{center}
\end{figure}

We consider the motion of two planets of mass $m_1$ and $m_2$ orbiting around a star of mass $m_0$ with their semi-major axes and mean longitudes $a_j$ and $\lambda_j$, respectively. When the semi-major axes of the two planets are close enough, in the quasi-coplanar quasi-circular case, the evolution of the resonant angle $\zeta=\lambda_1-\lambda_2$ can be modelled by the 2nd order differential equation \citep{Ed1977,RoNi2015}:
\begin{equation}
\ddot{\zeta}=-3 \eta^2 \mu \left( 1- (2- 2\cos \zeta)^{-3/2} \right) \sin \zeta\, ,
\label{eq:eqerdi}
\end{equation}
where $\mu=(m_1+m_2)/m_0$, and $\eta$ is the average mean-motion, {defined as the barycenter of the instantaneous mean-motion of the two planets: $(m_1 +m_2)\eta=m_1n_1  + m_2 n_2 $ \citep{RoRaCa2011}}. The phase space of eq. (\ref{eq:eqerdi}) is shown in Fig. \ref{fig:zeta}. $\zeta=180^\circ$ corresponds to the hyperbolic $L_3$ Lagrangian equilibrium, while $\zeta=\pm60^\circ$ are the stable configurations $L_4$ and $L_5$. Orbits that librate around these stable equilibria are called tadpole, or trojan (in reference to Jupiter's trojan swarms). Examples of trojan orbits are shown in purple in Fig. \ref{fig:zeta}. The separatrix emanating from $L_3$ (black curve) delimits trojan orbits from horseshoe orbits (examples are shown in orange), for which the system undergoes large librations that encompass $L_3$, $L_4$ and $L_5$. 

The libration of the resonant angle $\zeta$ is slow with respect to the average mean-motion $\eta$. The fundamental libration frequency $\nu$ is proportional to $\sqrt{\mu}\eta$. In the neighbourhood of the $L_4$ or $L_5$ equilibria, $\nu = \sqrt{27/4}\sqrt{\mu}\eta$  \citep{Charlier1906}. Away from the equilibrium, we compute $\nu$ by integrating eq. (\ref{eq:eqerdi}). Its value is given in Fig. \ref{fig:zeta} (lower panel) with respect to the initial value of the resonant angle.

\subsection{Stability of similar-period planets}
\label{sec:stab}

 \begin{figure}
\begin{center}
\includegraphics[width=1\linewidth]{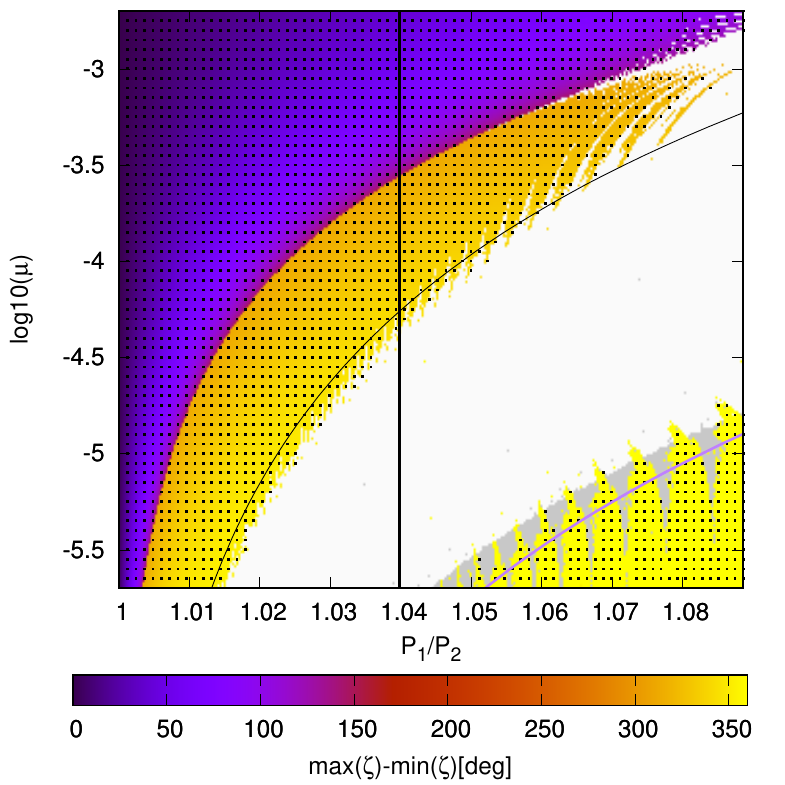}
\caption{\label{fig:stab_da} Stability map as function of $P_1/P_2$ and $\mu=(m_1+m_2)/m_0$, for circular coplanar orbits, taking $\zeta=60^\circ$ as initial condition (vertical dashed line in Fig \ref{fig:zeta}). Coloured pixels are long term stable with respect to the integration time of $10^5$ orbits. Grey pixels are long term unstable, and white pixels are short term unstable.  {The black curve represent the mutual Hill radius of the planets, $(P_1/P_2)^{2/3}=1+(\mu/3)^{1/3}$}, while the vertical line represents $P_1/P_2 \approx 1.04$ which is close the estimated value for the candidates of the TOI-178 system discussed in Sec. \ref{sec:TOI}. The color code is the amplitude of libration, black for zero (lagrangian equilibria), orange for horseshoe orbits. The bottom right corner shows stable orbits outside of the co-orbital area, in yellow. The purple line is the stability criterion from the overlapping of first-order mean-motion resonances \citep{DePaHo2013}. Black dots represent orbits that are stable over $10^{10}$ orbital periods, see the text for more details. }
\end{center}
\end{figure}

The main parameters we have access to when detecting a planet through transit surveys are the orbital period, the epoch of transit, the radius of the planet, and the impact parameter.
As the eccentricity of the planets are generally unconstrained by these observations, the stability of detected multi-planetary systems is estimated through criteria involving the mass and semi-major axis of the planets, such as Hill stability. Such criteria, however, does not take into account the stability domain of the coorbital 1:1 mean motion resonance. For coplanar circular orbits, the width of this domain (i.e. $a_1-a_2$ or $P_1-P_2$), is at its largest for $\zeta = \pm 60^\circ$, see Fig. \ref{fig:zeta}. As shown in previous studies \citep{LeRoCo2018}, the stability domains of these configurations depend mainly on $\mu=(m_1+m_2)/m_0$, and very little on $m_2/m_1$.

To illustrate the stability of circular coplanar co-orbitals, we integrate the 3-body problem for a grid of initial conditions. Taking initial conditions along the vertical dashed line on Fig. \ref{fig:zeta} ($\zeta(0)=60^\circ$) allows us to study all the possible co-orbital configurations in the coplanar circular case from the Lagrangian equilibria ($P_1/P_2=1$) to horseshoe orbits (stability around $L_5$ is obtained as well due to the symmetry of the problem). We hence integrate a grid of initial conditions along $P_1/P_2 \in [1,1.09]$ for various values of $\mu= (m_1+m_2)/m_0 \in [2\times 10^{-6 },2\times 10^{-3}]$. The results of these integrations are shown in Fig. \ref{fig:stab_da}.

For each set of initial conditions, the system is integrated over $10^{5}$~orbital periods using the symplectic integrator SABA4 \citep{LaRo2001} with a time step of $0.01001$ orbital periods. Trajectories with a relative variation of the total energy above $10^{-7}$ are considered unstable.
Such trajectories are identified with white pixels. These short term instabilities are generally due either to the overlap of secondary resonances in the co-orbital region \citep[][]{RoGa2006,PaEf2015, PaEf2018}, or to the overlap of first-order mean motion resonances outside this domain \citep{Wi1980,DePaHo2013,PeLaBo2017}. 
Grey pixels identify the initial conditions for which the diffusion of the mean motion of one of the planets between the first and second half of the integration is higher than  $10^{-5.5}$ \citep{Laskar1990,Laskar1993,RoLa2001}.
The integration time of $10^5$ orbital period is not enough to assess the stability on the lifetime of a planetary system. However, from the estimates regarding the diffusion variation versus time given in \cite{RoLa2001} and \cite{PeLaBo2018}, we deduce that a mean-motion diffusion rate lower than $10^{-7}$ derived from integrations over $10^{7}$ orbits enable us to ensure the stability over $10^{10}$ orbits. This was checked for a lower resolution grid of initial conditions. Giga-year stable orbits are shown by black dots on Fig. \ref{fig:stab_da}.

The color code represents the libration amplitude of the resonant angle $\zeta$: purple for trojan orbits, orange for horseshoe, and yellow if the configuration is outside the co-orbital resonance but on a stable orbit. {Due to the chosen resolution of initial conditions, the chaotic area in the vicinity of the separatrix between the trojan and horseshoe domains is visible only for large masses.}  
 The purple line represents the stability criterion proposed by \cite{DePaHo2013} for the outer limit of the
chaotic area:  $(P_1/P_2)^{2/3} =1+1.46 \mu^{2/7}$.
The black line indicates $P_1/P_2\approx 1.04$, which is close to the estimated value for both TOI-178 and Kepler-132 candidates. Pairs of planets have to be either above the black curve (in the co-orbital resonance), or bellow the purple one (on separated orbits), to be on long-term stable orbits.

\subsection{Transit Timing Variations of similar-period planets}
\label{sec:TTV}

%
%
%

Using the notations, {reference frame}, and results of \cite{LeRoCoLi2017}, planet $j$ transits, at first order in eccentricities {and inclinations}, when 
\be
\lambda_j=-\pi/2+2 e_j \cos (\varpi_j)\, ,
\label{eq:transitj}
\ee
where 
\be
\begin{aligned}
\lambda_1(t) & =\lambda_0 + \eta t + \frac{m_2}{m_1+m_2} \zeta(t) \, , \\
\lambda_2(t) & =\lambda_0 + \eta t - \frac{m_1}{m_1+m_2} \zeta(t) \, ,
\end{aligned}
\label{eq:TTVg}
\ee
and $\zeta(t)$ is given by eq. (\ref{eq:eqerdi}). {The TTVs induced by the co-orbital configuration are detailed in \cite{FoHo2007} and \cite{VoNe2014}. In this study we simply comment on the following point: 
a pair of co-orbital exoplanets might be mistaken for two planets on close but distinct non-resonant orbits if:\\
\noindent - Their libration period is significantly longer than the duration of the observation. In this case, the continuous evolution of the instantaneous period of the co-orbitals can be retrieved using eq. (\ref{eq:TTVg}). This is the case of the TOI-178 that will be discussed in section \ref{sec:TOI}.\\ 
\noindent - They are on horseshoe orbits and have similar radii. This case is discussed in the rest of this section.\\
In both cases, since the orbits are within the Hill instability region, it might lead to the rejection of one of the candidate as false positive despite the fact that the planets are on stable co-orbital orbits. }\\

 \begin{figure}
\begin{center}
\includegraphics[width=1\linewidth]{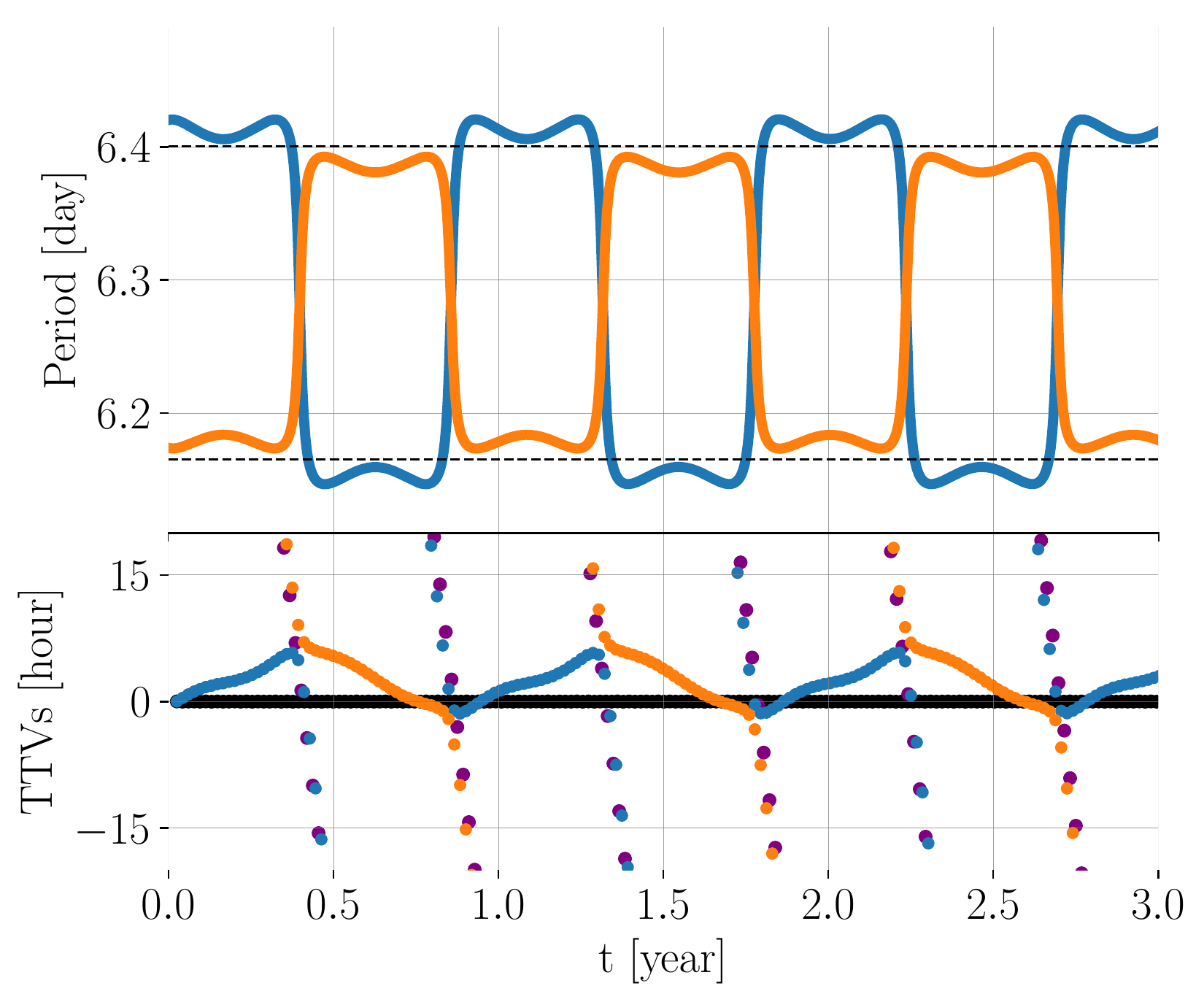}
\caption{\label{fig:TTV_KOI} Top panel: instantaneous orbital period of two planets in a horseshoe orbit.   Bottom: TTVs river diagram of the transits of both planet with respect to the outer dashed line at $\sim 6.4$day. In the background, black (resp. purple) dots give the river diagram of a planet in the isolated orbit $P_1$ (resp. $P_2$). }
\end{center}
\end{figure}

We illustrate the horseshoe orbit case using the Kepler-132 system which has four validated planets \citep{Rowe2014}, two of them with $P_1=6.4149$ day and $P_2=  6.1782$ day (planets Kepler-132\,c and Kepler-132\,b). As the central star of Kepler-132 was shown to be a possible wide binary, these 2 planets were subsequently announced to orbit each a different component of the binary, as they would be Hill-unstable if they orbited the same star \citep[]{Lissauer2014}. We call this scenario (i). {In this case, no significant TTVs are expected for these two planets.}

{We propose an alternative scenario (ii) that we illustrate in Fig. \ref{fig:TTV_KOI}. For this figure, we integrated the orbit of two masses $m_x=3\times 10^{-5} m_0$, $m_y=3.75\times 10^{-5} m_0$ around a $ m_0=1.37$ solar mass star, in a horseshoe orbit. Two planets in such orbit `exchange' their position every half libration period. We approximate this motion by `jumps' between an upper period: $P_{x,+}$ (resp. $P_{y,+}$), and a lower period  $P_{x,-}$ (resp. $P_{y,-}$) for the planet $x$ (resp. $y$). 
} 

{For this example, we chose initial conditions for the planet $x$ and $y$ such that the average of the upper positions $(P_{x,+}+P_{y,+})/2$ (resp. lower position $(P_{x,-}+P_{y,-})/2$)  are the orbital periods announced for Kepler-132\,c (resp. Kepler-132\,b), shown in black dashed lines. As the swapping between higher and lower period for $x$ and $y$ happen when they are near conjunction, and is quick with respect to the libration time scale, both scenario (i) (planet 1 and 2) and (ii) (planet $x$ and $y$) yield similar transit timings: the bottom panel of that figure represents the simulated river diagram of that system, folding the time of transits of the planets $x$ and $y$ with respect to the fixed outer period $P_1$. If we assume that we cannot distinguish the transits of planet $x$ and $y$, they can be mistaken for planet $1$ having moderate libration around the orbit of period $P_1$ (black dots in bottom panel of Fig \ref{fig:TTV_KOI}), and planet $2$ oscillating around a distinct orbit $P_2$ (purple dots), which result in almost-vertical lines in this river diagram.
} 

{We hence consider the possibility that the announced planets 1 and 2 of the Kepler system would instead be the planets $x$ and $y$ represented in blue and orange in Fig. \ref{fig:TTV_KOI}. In this case planets 1 and 2 would be `fictitious' planets that are alternately planet $x$ and $y$: the `planet 1' which has an announced period of $P_1=6.4$ day is actually half of the time the planet $x$, and the rest of the time the planet $y$, and the same goes for planet 2. This scenario is possible because the two announced planets have nearly indistinguishable radii ($R_1=1.3 \pm 0.3$ and $R_2=1.2 \pm 0.2$). }


However in the scenario (ii) the fictitious planets 1 and 2 exhibit significant TTVs when the actual planets $x$ and $y$ have different masses (see bottom panel of Fig. \ref{fig:TTV_KOI}): the instantaneous period of each planet librates around a {mean orbital period $\overline P = (P_1+P_2)/2= (m_xP_x+m_yP_y)/(m_x+m_y)$ \citep{RoRaCa2011}}. The upper and lower position of the planets $x$ and $y$ read:
\be 
P_{x,\pm}= \overline P \pm \frac{m_y \delta P}{m_x+m_y}, \ \ P_{y,\pm}= \overline P \pm \frac{m_x \delta P}{m_x+m_y},
\ee 
with $\delta P= (P_1-P_2)$. This jump is occurring at every conjunction,
$P_{\text{swap}} = (1/P_2-1/P_1)^{-1} $. As $P_1$ is the averaged value of $P_{x,+}$ and $P_{y,+}$, the TTVs have an amplitude of:
\be
TTVs=(P_{x,+}-P_1)P_{\text{swap}}/\overline P = \frac{m_y-m_x}{m_x+m_y} \frac{P_1P_2}{P_1+P_2}\, ,
\label{eq:TTVkoi}
\ee
{ and a period of $2P_{\text{swap}}$. Note that the amplitude is sightly overestimated as the instantaneous periods are continuously swapping instead of jumping. This smooth evolution also produces TTVs but they are negligible with respect to those described by eq. (\ref{eq:TTVkoi}) as long as $m_x$ and $m_y$ differ by more than a few percent. }
 {In Fig. \ref{fig:TTV_KOI}, $m_x/m_y=5/4$, inducing TTVs estimated at 8.37 hour by eq. (\ref{eq:TTVkoi}). }
 
{In the Kepler-132 case, our light-curve analysis excluded TTVs larger than half an hour on both planets 1 and 2. If we consider the scenario (ii), i.e. that the transits are instead produced by two planets $x$ and $y$ on horseshoe orbits, eq. (\ref{eq:TTVkoi}) yields $m_x-m_y<0.0066 (m_x+m_y)$, implying a mass difference below $1.5\%$ between the two planets. Albeit the scenario (i), where each star of a binary has a similar planet at almost equal orbital periods seems unlikely, the scenario (ii), where two planets on a horseshoe orbit have equal masses down to the percent level doesn't seem much likelier.}




%

\section{TOI-178}
\label{sec:TOI}

%
\begin{table}
\caption{Parameters for the 3 candidates in the TOI-178 system, extracted from the data validation report (DVR) of the TESS mission.}
\label{table:TOI}
\centering
\setlength{\extrarowheight}{2pt}
\begin{tabular}{l c c c}
\hline\hline
Parameter &  value  \\
\hline
\multicolumn{2}{c}{Star (TOI-178)}\\
$m_0$ [$ M_{sun}$]& $0.643 \pm 0.075$\\
$R_0$  [$ R_{sun}$] & $0.70 \pm 0.15$ \\
\hline
\multicolumn{2}{c}{Planet 1 (TOI-178.02)}\\
$P_1$ [day] & $10.3542 \pm 0.0032$ \\
$T_1$ [BTJD] & $1354.5522\pm 0.0041$ \\
$R_1$ [$R_e$] & $3.7\pm 1.5$\\
\hline
\multicolumn{2}{c}{Planet 2 (TOI-178.03)}\\
$P_2$ [day] & $9.9559\pm 0.0051$\\
$T_2$ [BTJD] & $1362.9533 \pm 0.0035$   \\
$R_2$ [$R_e$] & $2.3 \pm 2.7$ \\
\hline
\multicolumn{2}{c}{Planet 3 (TOI-178.01)}\\
$P_3$ [day] & $6.5581\pm0.0013$\\
$T_3$ [BTJD] & $1360.2423\pm 0.0024$ \\
$R_3$ [$R_e$]  & $2.8\pm 1.1$ \\
   \hline   \hline
\end{tabular}
\end{table}

The first release of candidates from the TESS alerts of Sector 2 included three planet candidates in the TESS Object of Interest (TOI) TOI-178 (or TYC 6991-00475-1). The candidates TOI-178.01, TOI-178.02 and TOI-178.03 transited respectively 4, 3 and 2 times during the 27 days of observation. The TESS pipeline fits converge on a solution where all three candidates are of planetary nature. In the TESS data validation report all three planets pass all first order tests to exclude potential false positive signals. 
The in difference image centroid offsets relative to the TESS Input Catalog position and relative to the out of transit centroid are within 2 sigma for all 3 planet candidates (except 2.09 sigma for planet 3), showing no indication for an eclipsing binary scenario.
The pipeline ghost diagnostic test that searches for correlation of time series of core and halo apertures ruled out optical ghosts of bright eclipsing binaries outside of the target apertures as the source of the transit-like features. The bootstrap test excludes a false alarm scenario by $10^{-13}$ or smaller for all three planetary candidates. The only two “yellow” flags are the 7.6 SNR and slight centroid offset of 2.09 sigma for the candidate TOI-178.03.\\

{During the TESS observation of sector 2, the period ratios of the two outer candidates was close to 1.04 (see Tab. \ref{table:TOI}), while the phase between these planets ranged between $\zeta \in [280^\circ:310^\circ]$ (assuming circular orbits), which is in the vicinity of the $L_5$ equilibria ($\zeta=300^\circ$, see Fig. \ref{fig:zeta}). Besides, the radius of the outermost candidate is estimated to be  $3.7\pm 1.5$ Earth radii, while the radius of the other candidate is poorly constrained. An estimation of the mass of these candidates is difficult, albeit the outermost appears to be a `Neptune-like' object \citep{CheKi2017}, and is hence very unlikely to have a sub-Earth mass. As a result, as shown in Sec. \ref{sec:stab}, these candidates have to be in the co-orbital resonance in order to be on stable orbits, see the vertical line in Fig. \ref{fig:stab_da}. This prompted us to study this system in more details. For consistency with the section \ref{sec:general}, in the rest of the paper we name 1 and 2 the two co-orbital candidates, while the planet at 6.5 day will be called planet 3, see Tab. \ref{table:TOI}.}


\subsection{Independent transit search and orbital fit}
\label{sec:TOIlc}

We run an independent transit search to confirm the result of the TESS Science Processing Operations Center (SPOC) pipeline \citep{jenkins2016}, starting from the target pixel file, which we downloaded from the MAST\footnote{Mikulski Archive for Space Telescopes, https://archive.stsci.edu/prepds/tess-data-alerts/.}. We first calculate the centroid position and  Full Width at Half Maximum (FWHM) of the target's point-spread-function (PSF) for each frame. We then generate the lightcurve by using a circular top-hat aperture, tracking the PSF center in each frame.

Our transit search and detrending pipeline is based on the Gaussian Process (GP) pipeline used to detrend K2 lightcurves in \cite{luger2017,grimm2018}, modified to work with the higher cadence rate of TESS lightcurves. We find that the systematic noise encountered in TESS lightcurves has a different source than in K2. Namely, instead of being correlated with the PSF centroid offset, it has a strong correlation with the x and y FWHM of the target PSF. Therefore, we employ a similar procedure as in  \cite{luger2017}, running a GP regression pipeline to simultaneously fit the systematic noise correlated with the x and y FWHM, and the longer-term fluctuations caused by stellar variability. The noise is then subtracted to produce a flattened lightcurve. The detrended lightcurve has a similar residual noise level as the
PDCSAP lightcurve produced by the SPOC pipeline (with median absolute deviations of 0.126\% and 0.124\% respectively).

This GP-based detrending method is independent from the TESS pipeline, which uses co-trending basis vectors derived from the entire ensemble of lightcurves in a dataset \citep{jenkins2016}. As a result, we can provide an independent check that the transit signals are not spuriously caused by data processing.

We run a standard transit search to find the 10 strongest transit-like signals, based on similar procedures in e.g. \cite{vanderburg2016,dressing2017,mayo2018}. First, we perform a series of BLS fits to find periodic dimming events in the lightcurve, removing each signal for subsequent fits \citep{kovacs2002}. For each of the 10 signals, we fit a transit model based on \cite{mandel2002}. We use the batman package to compute the transit model \citep{kreidberg2015}, inputing limb-darkening parameters calculated from \cite{claret2011}.

Running the transit search on our detrended lightcurve, we recover all three transit signals found by the TESS pipeline. However, we also find a strong signal with a 4.96 day period, and a first-transit time of $t_0 = 1458.006$ (BTJD). This is an alias of candidate 2. We calculate the log-likelihood difference between the 9.96 period and its alias to be 7.74, in favour of the 9.96 day period. This would correspond to a difference of $\sim -15$ in the Bayesian Information Criterion, which is considered significant \citep{schwarz1978}. However, our limited constraints on the properties of the host star will have an effect on the likelihood ratio, as it is calculated based on a transit model which requires stellar mass and limb-darkening. \\

\begin{table}
\caption{Orbital fit to the transit timings observed by TESS. The initial conditions are taken close to the 3:2:2 resonant chain in scenario (i), and close to the 4:3:2 resonant chain in scenario (ii). The eccentricities were set with an upper limit at 0.2, and were unconstrained by the fit.}
\label{table:M}
\centering
\setlength{\extrarowheight}{4pt}
\begin{tabular}{l c c c}
\hline\hline
Parameter & value (i) & value (ii)\\
\multicolumn{3}{c}{Planet 1 (TOI-178.02)}\\
 $P_1$ [day]  & $ 10.2601_{-0.111}^{+0.107}$& $10.2974_{-0.110}^{+0.053}$ \\
  $T_1$ [BTJD] & $ 1354.57_{-0.04}^{+0.05}$& $1354.56_{-0.02}^{+0.03} $ \\
   $m_1/m_0$[1E-4] & $5.77_{-4.01}^{+7.28}$& $4.35_{-3.66}^{+7.79}$ \\
\hline       
\multicolumn{3}{c}{Planet 2 (TOI-178.03)}\\
 $P_2$ [day] & $ 9.9766_{-0.117}^{+0.188}$& $4.9032_{-0.155}^{+0.067}$ \\
  $T_2$ [BTJD]  & $1353.02_{-0.09}^{+0.11}$& $1348.07_{-0.12}^{+0.13}$ \\
  $m_2/m_0$[1E-4]&  $3.11_{-2.47}^{+3.79}$ & $0.73_{-0.67}^{+2.34} $\\
   \hline       
\multicolumn{3}{c}{Planet 3 (TOI-178.01)}\\
  $P_3$ [day]  & $ 6.6053_{-0.048}^{+0.099}$& $6.60771_{-0.044}^{+0.095}$ \\
   $T_3$ [BTJD]  & $1353.66_{-0.04}^{+0.03}$& $1353.68_{-0.03}^{+0.03}$ \\
    $m_3/m_0$[1E-4] & $ 5.66_{-4.60}^{+5.54}$& $3.68_{-2.85}^{+5.81}$ \\
\hline 
\hline        
\end{tabular}
\end{table}

%

{Given the signal to noise ratio limit of the TESS observations, we cannot fully confirm the 9.96d period of candidate 2 based on individual transits. We then check if the transit timing variations can help us discriminate between the two scenarios: (i) The configuration announced by the TESS pipeline, where the orbits are near a 3:2:2 resonant chain; and (ii) The case where the candidate 2 is in a 4.96 day period, resulting in a near 4:3:2 resonant chain. The date of individual transits, derived both from the lightcurve detrended by the TESS pipeline and the result of our own detrending, are given in Tab. \ref{tab:posterior_ttvs}.}

{We perform the Transit Timing Variation (TTV) analysis with an ensemble differential evolution Markov chain Monte Carlo method (DEMCMC) \citep{Braak2006,Vrugt2009}, similar as described in \cite{grimm2018}. This method is using the GPU N-body code GENGA \citep{Grimm2014} to calculate the orbital evolution of the planets and the transit times for the DEMCMC steps. The estimated masses, in each scenario, are summarised in Tab. \ref{table:M}, while the posterior distribution functions can be found in Fig. \ref{fig:MA1}. We note that these posteriors do not take into account the long-term stability of the fitted orbits. Due to the low number of observed transits, and the short time span of the observations, the uncertainties on the masses are quite large, and do not allow to discriminate further a scenario with respect to the other. We note however a difference between the posterior distributions of $m_2/m_0$ between the two scenarios: if the candidate 2 is on a $4.94$ day orbit, then its mass should be significantly smaller than the other two candidates. }\\

{Our analysis of the current observations does not allow to fully discard the 4:3:2 scenario. In the next sections we nonetheless consider the case that is favoured by both our lightcurve analysis and the TESS pipeline: the near 3:2:2 resonant chain. The stability of the orbit of planets 1 and 2 was already discussed in section \ref{sec:stab}. However, the presence of the planet 3 near a 2:3 MMR with the co-orbital pair might further reduce their stability domain \citep{RoGa2006}. We analyse the stability of this 3-planet system in the next section.}



\subsection{Stability analysis}

 \begin{figure}
\begin{center}
\includegraphics[width=1\linewidth]{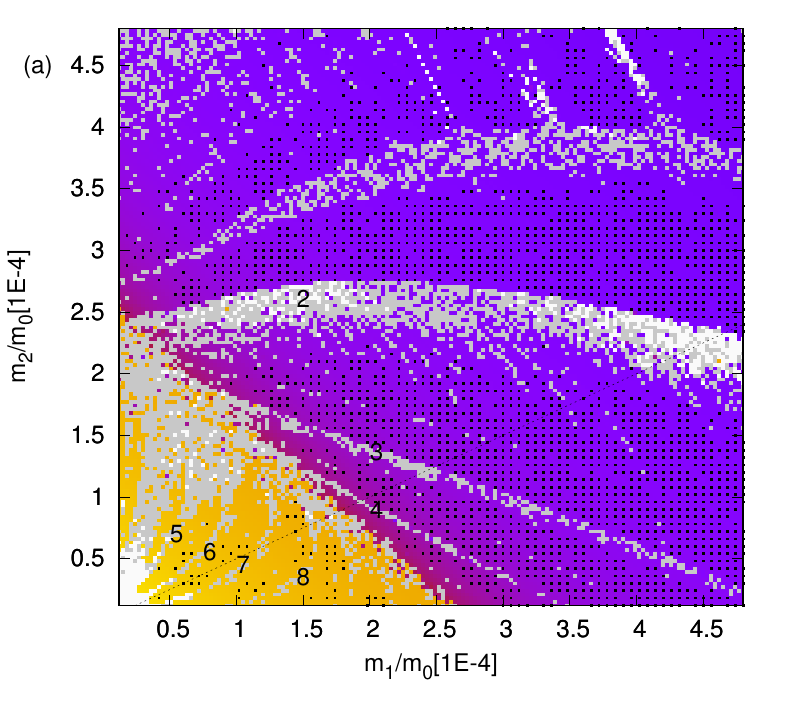}\\
\vspace{-.3cm}
\includegraphics[width=1\linewidth]{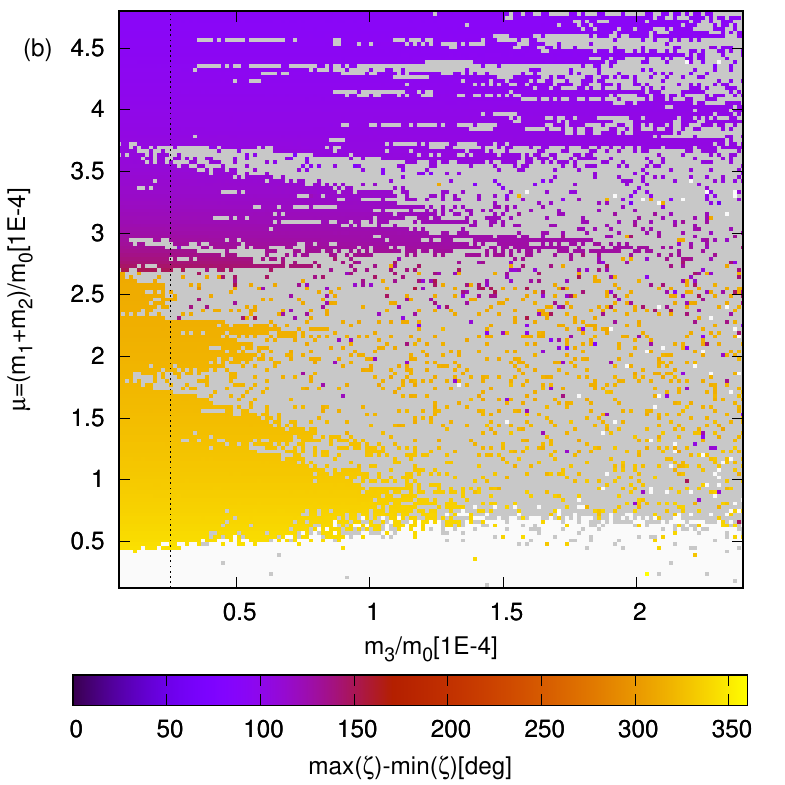}\\
\caption{\label{fig:stab_TOI2} Stability domains for the co-orbital candidates of the TOI-178 system. (a) as a function of the mass of the two co-orbitals, fixing $m_3=2.5\times 10^{-5}m_0$; and (b) as a function of the mass of the two coorbitals $\mu$ and the inner planet $m_3$. Dotted lines represent the intersection of the planes of initial conditions (a) and (b). The color code is the same as Fig. \ref{fig:stab_da}. Panel (a) shows black dots to represent orbits that are stable over $10^{10}$ orbital periods, see Sec. \ref{sec:stab} for more details (orbit for higher masses are currently being integrated and will be displayed in the re-submitted version). The numbers displayed in panel (a) are the value of $p$ for the disruptive $\Psi = p \nu +g $ resonances.}
\end{center}
\end{figure}
%
%

As $P_1/P_3 \approx 1.58$ and $P_2/P_3 \approx 1.52$, the co-orbital configuration is just outside the 3:2 MMR with planet 3. As a result, we expect resonances between the libration frequency $\nu$ of the co-orbitals and the frequency $\Psi=2n_3-3\eta$ (called great inequality), where $n_3$ is the mean-motion of planet 3 and $\eta$ the average mean-motion of the co-orbitals \citep{RoGa2006}. As we have no information on the eccentricities of the planets, we analyse the stability in the circular case. Both $\nu$ and $\Psi$ depend on the averaged mean-motion, which in turn depends on the mass of the star and the mass ratio between the two co-orbitals. The mass of the star is a common factor to all involved frequencies and can hence be ignored, as the initial conditions are taken using the relative orbital periods of the planets. Given the constraints we have on the orbits, the three main parameters for the stability of the system are the masses of the three planets.



We hence check the stability of the co-orbital configuration as a function of $m_1/m_0$, $m_2/m_0$, and $m_3/m_0$ in two different planes: in Fig. \ref{fig:stab_TOI2} (a) we vary $m_1/m_0$ and $m_2/m_0$, taking an arbitrary value for $m_3=2.5\times 10^{-5} m_0$($\approx 5.3 M_{earth}$, using $m_0$ given in Table \ref{table:TOI}); while in (b) we fix $m_1/m_2$  varying $\mu$ and  $m_3/m_0$. The initial values of the mean longitudes and semi-major axis are derived from Table \ref{table:TOI}, and the orbits are initially coplanar and circular. These figures are obtained in the same way and have the same color code as in Fig. \ref{fig:stab_da}, see section \ref{sec:stab}.

In Fig. \ref{fig:stab_TOI2} (a), we recover the stable domains for horseshoe and trojan orbits, as was the case of Fig. \ref{fig:stab_da}. We see however that these stability domains are crossed by disruptive resonances. We identified the main chaotic structures to be in the wake of resonances of the form $\Psi = p \nu +g $, where $g$ is one of the three secular frequencies of the system, small with respect to $\nu$. The $p=[2,3,4]$ resonances cross the trojan area, while $p=[ 5, 6, 7, 8 ]$ disturb the horseshoe domain \citep[for more details, see][]{RoGa2006}. For two isolated co-orbitals $m_1/m_2$ has close to no impact on the stability of the orbits in the coplanar quasi-circular case \citep{LeRoCo2018}. Here however, the mass repartition between the co-orbitals shift the value of the averaged mean-motion $\eta$, displacing the positions of the disruptive resonances. This effect, combined with the evolution of the resonant frequency $\nu$ (which is function of both $\mu$ and the amplitude of libration of $\zeta$, see Fig \ref{fig:zeta}), gives the unstable structures displayed in Fig. \ref{fig:stab_TOI2}. Long term-stable areas remain for many values of $m_1$ and $m_2$, but $m_1>m_2$ is overall favoured by this stability analysis.

Fixing the mass ratio to an arbitrary value $m_1/m_2=2$, and changing $\mu$ and $m_3/m_0$, we show in Fig. \ref{fig:stab_TOI2} (b) that $m_3/m_0$ has little effect on the position of the resonant structures, as it does not impact the value of the concerned frequencies beside the secular frequency $g$. As a result, an increase of $m_3/m_0$ only increases the width of the chaotic area near these resonances, further reducing the stability domains. A good estimation of the mass of the planet 3 can hence further constrain the possible co-orbital configuration of the planets 1 and 2. It is important to bear in mind, however, that panel (b) only represents the evolution of the $m_1=2m_2$ line on panel (a), and hence a more detailed analysis is required once more constrains are obtained on the masses.


\subsection{Predicted TTVs for future observations}

\begin{figure}
\begin{center}
\includegraphics[width=1\linewidth]{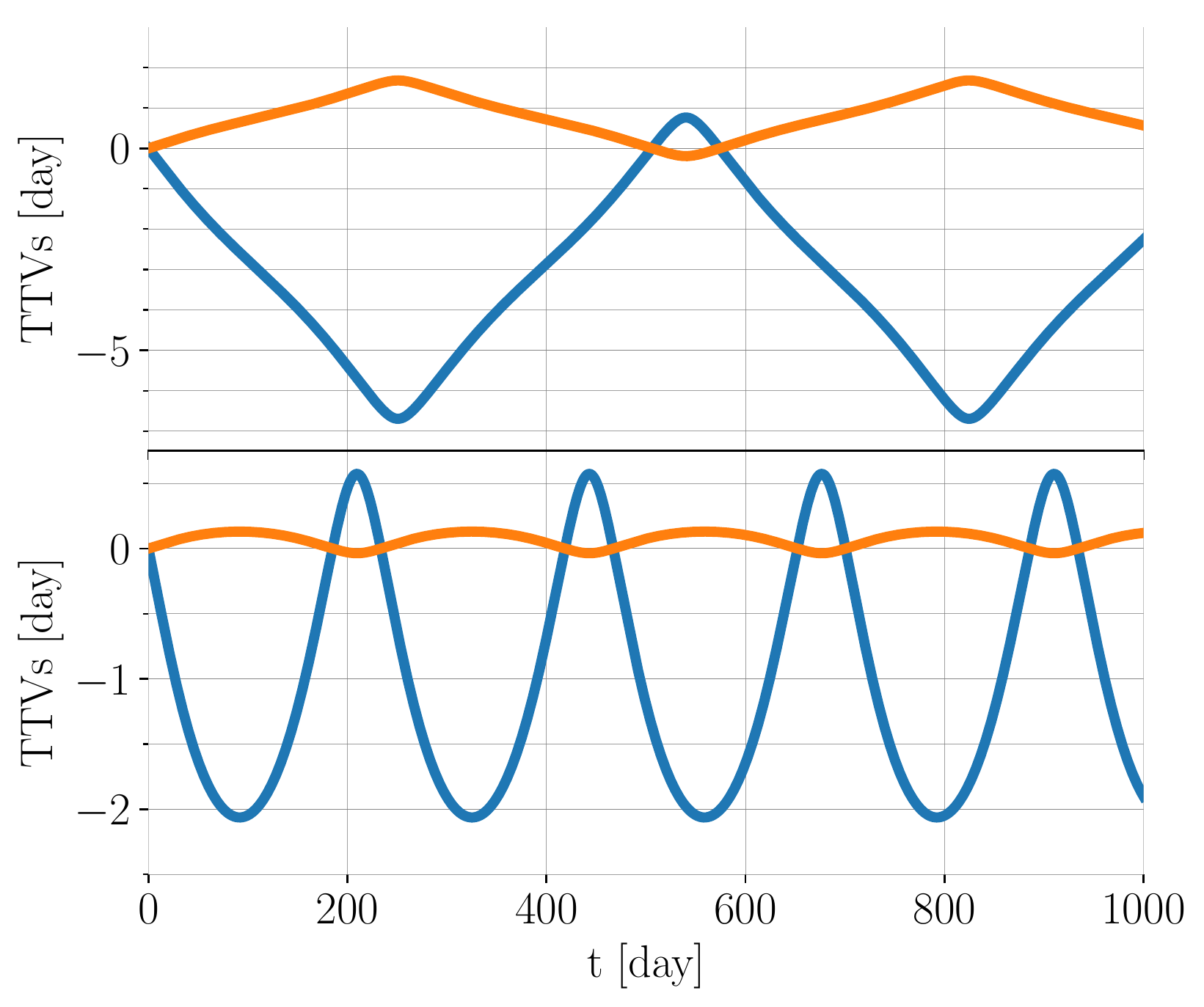}
\caption{\label{fig:TTV_ex} Example of TTVs for TOI-178.02 ($m_1$, orange) and TOI-178.03 ($m_2$, blue) for two arbitrary sets of masses, taking as initial conditions the configuration of the system during the observation of sector 2 by TESS. Top panel: $m_1=1 \times 10^{-4} m_0$, $m_2=2.5 \times 10^{-5} m_0$, resulting in a horseshoe configuration. Bottom panel: $m_1=4 \times 10^{-4} m_0$, $m_2=2.5 \times 10^{-5} m_0$, resulting in a large amplitude trojan configuration. }
\end{center}
\end{figure}

\begin{figure}
\begin{center}
\includegraphics[width=1\linewidth]{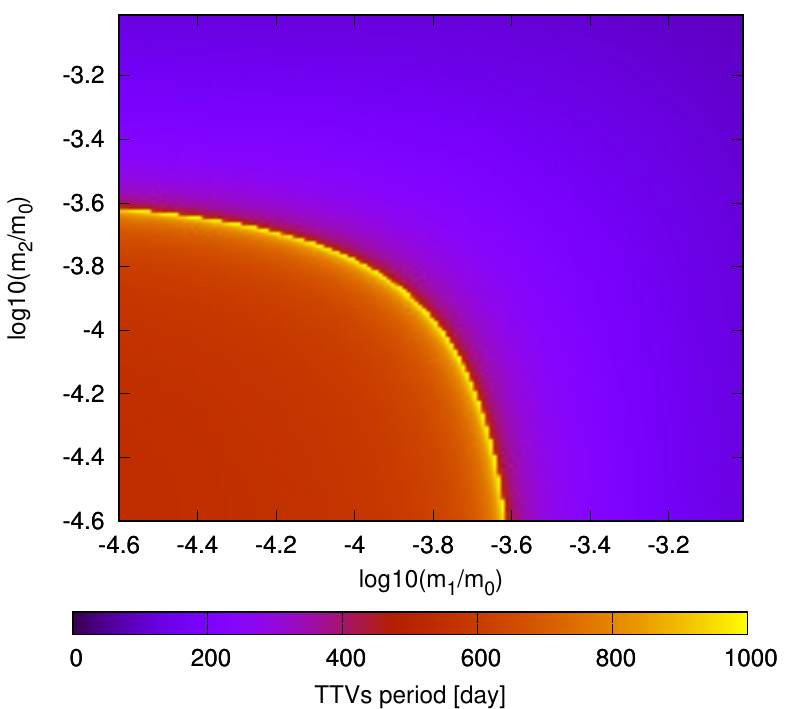}\\
\includegraphics[width=1\linewidth]{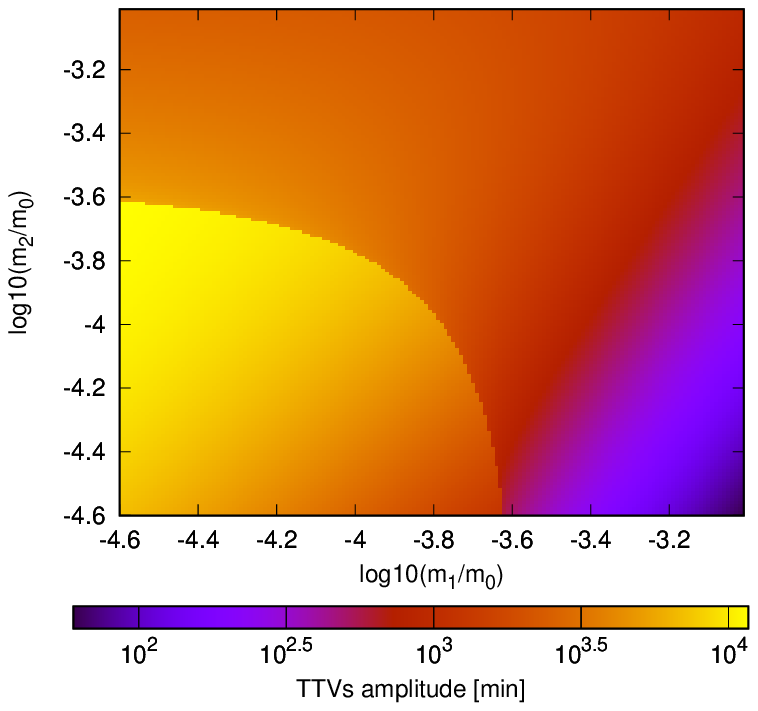}\\
\caption{\label{fig:TTV_TOI} Period (top) and amplitude (here for planet 1, TOI-178.02, bottom) of the predicted TTVs induced by the co-orbital motion, assuming circular orbits derived from Table \ref{table:TOI}. The effect of planet 3 is neglected. The amplitude of TTVs predicted for planet 2 (TOI-178.03) are obtained by swapping the x and y labels of the bottom panel.}
\end{center}
\end{figure}

{Depending on $m_1$ and $m_2$, we showed in the previous section that the TOI-178 system could harbour stable co-orbital exoplanets, either in a trojan or horseshoe configurations. Using equation (\ref{eq:TTVg}), we show in Fig. \ref{fig:TTV_ex} the TTVs that should exhibit such configurations, taking as initial conditions the orbital elements summarised in Tab. \ref{table:TOI}, along with two set of arbitrary masses. In the top panel, $m_1+m_2= 1.25 \times 10^{-4} m_0$, resulting in a horseshoe orbit (see Fig. \ref{fig:stab_da}), while in the bottom planet $m_1+m_2= 4.25 \times10^{-4} m_0$, which result in a tadpole orbit with a large amplitude of libration.}

{Fig. \ref{fig:TTV_ex} also illustrates why planet 1, despite transiting 3 times during the observation by TESS of the sector 2, would not display significant TTVs during that time: in both example, the evolution of the TTVs is quasi-linear over the first 30 days. As a result, these TTVs can be absorbed by a redefinition of the orbital period of the planet. This linearity is due to the fact that $\zeta \in [280^\circ:310^\circ]$ during the observations, which correspond to a global extremum of the instantaneous period of both co-orbitals regardless of their amplitude of libration, see Fig. \ref{fig:zeta}.}\\

{For this system, the full TTVs induced by the co-orbital motion happen over hundreds of days. Figure \ref{fig:TTV_TOI} gives the amplitude and period of the TTVs expected for planet 1 for a grid of masses of the co-orbital candidates. The effect of the inner planet 3 is neglected, and should be small compared to the libration in the co-orbital resonance. The TTVs' amplitude of each planet is proportional to the mass of the other planet, and proportional to the resonant angle (eq. \ref{eq:TTVg}). As a result, the amplitude of the TTVs expected for planet 2 can easily be deduced by swapping the labels of the x and y axis in the lower panel of Fig. \ref{fig:TTV_TOI}.}

{Due to the evolution of the resonant angle of $\approx 30^\circ$ during the observation of sector 2 by TESS, at least one of the two co-orbital candidates should exhibit TTVs of the order of a day or more. If these TTVs are detected, it would not only confirm the existence of the co-orbital pair, but also allows for unique and precise determinations of $m_1/m_0$ and $m_2/m_0$. }

\section{Summary and Conclusions}

In section \ref{sec:general} we have reviewed the main properties of the 1:1 mean-motion resonance in the case where both objects are of planetary nature, and we gave the constrains on TTVs that allows us to recover if two planets on apparently close period orbits are actually in a horseshoe configuration. We applied this method to the Kepler-132 system, {and concluded that Kepler-132\,b and Kepler-132\,c need to have equal masses down to the percent level for their TTVs to be consistent with the co-orbital hypothesis.}\\

In section \ref{sec:TOI} we have analysed in detail the case of the TOI-178 system, where two planet candidates appear to be on a co-orbital orbit. All first order analysis described in the TESS data validation report point at transit signals of planetary nature coming from one system in TOI-178. One question still to be addressed is if these detected transits were correctly accounted to individual planets. Our independent detrending and transit search recovers all three transit signals found by the TESS pipeline at the 6.5, 9.9, and 10.5 day periods, but we cannot exclude an alias for the TOI-178.03 (planet 2 in our study) which would put that planet on a 4.9 day orbit, resulting in a configuration in, or close to, a 4:3:2 three-body resonance. {Our TTVs analysis showed however that in this case, TOI-178.03 is expected to be significantly less massive than the other two candidates.}
A potential other alternative scenario would be that the two transits of the TOI-178.03 are indeed two individual transits of outer planets.
%
%

Assuming that the orbits summarised in Tab. \ref{table:TOI} are correct, we performed a stability analysis of the system allowing for a large range of mass for each planets. As long as the mass of the inner planet is not much more massive than the sum of the mass of the two co-orbital candidates, stable co-orbital configurations can exist for billion years. The stability analysis favoured the case where the TOI-178.02 is more massive than TOI-178.03.
During the time of the observation of the sector 2 by TESS, the phase between the two candidates was $\lambda_1-\lambda_2 \in[280^\circ,310^\circ]$. This allows the bodies to be on a vast range of amplitudes of libration around the $L_5$ Lagrangian point ($\zeta=300^\circ$) or in a horseshoe orbit (see Fig. \ref{fig:zeta}), depending on their mass. The minimal values of the mass of the bodies are also constrained by the stability diagrams Fig. \ref{fig:stab_da} and \ref{fig:stab_TOI2}.  

TTVs that should be induced by the co-orbital motion during the three transit observed of the TOI-178.03 cannot be used to further constrain the system because $\zeta \in[280^\circ,310^\circ]$ correspond to a local extremum for the instantaneous period of the bodies. Important long term TTVs, on an observation timespan of hundred of days, should however be observed on at least one of the the two candidates, see Fig. \ref{fig:TTV_ex} and \ref{fig:TTV_TOI}. 
Such TTVs must be observed if the orbits reported in Table \ref{table:TOI} are correct, and would allow to constrain the masses of the TOI-178.02 and TOI-178.03 with great precision.

Alternatively radial velocity measurements can be used, both on their own \citep{LauCha2002,LeRoCo2015}, and in combination with the transit measurements \citep{FoGa2006,LeRoCoLi2017}, to confirm the co-orbital nature of the system.
More constrains on TTVs might be provided by GAIA \citep{GAIA}, and by the ESA mission CHEOPS to be launched in fall 2019 \citep{CHEOPS}.

\begin{acknowledgements}
The authors acknowledge support from the Swiss NCCR PlanetS and the Swiss National Science Foundation. We thank Juan Cabrera for useful discussions. J.Schneider is grateful to Françoise Roques for preliminary discussions. A. Correia acknowledges support from projects UID/FIS/04564/2019, POCI-01-0145-FEDER-022217, and POCI-01-0145-FEDER-029932, funded by COMPETE 2020 and FCT, Portugal.
\end{acknowledgements}

\bibliographystyle{aa}
\bibliography{biblio}

\appendix

\section{TTVs of the TOI-178}
\label{ap:TTV}

\begin{table}[]
\setlength{\extrarowheight}{3pt}
\caption{Posterior of the transit times for the individual transits of each of the three planet candidates in TOI-178, both from the light curve detrended by the TESS pipeline, and our own analysis. All dates agree within $1\sigma$.}
\begin{center}
\begin{tabular}{lccccccccc}
\hline
\hline
Candidate &  ID & $T$ [day]  & $T$ [day] \\
 & transit &  TESS detrending & Our detrending \\
 \hline
Candidate 1  & 1 & $1354.5500_{-0.0058}^{+0.0110}$  &			
$1354.5550_{-0.0059}^{+0.0056}$ \\
Candidate 1  & 2 & $1364.9110_{-0.0190}^{+0.0190}$    & 
$1364.9089_{-0.0079}^{+0.0069}$ \\
Candidate 1  & 3 & $1375.2592_{-0.0060}^{+0.0045}$ & $1375.2598_{-0.0021}^{+0.0024}$ \\
Candidate 2  & 1 & $1362.9540_{-0.032}^{+0.033}$ &$1362.9485_{-0.0101}^{+0.0094}$ \\
Candidate 2  & 2 & $1372.9090_{-0.018}^{+0.014}$  &$1372.9106_{-0.0120}^{+0.0070}$ \\
Candidate 3  & 1 & $1360.2384_{-0.0042}^{+0.0045}$ & $1360.2381_{-0.0042}^{+0.0040}$\\
Candidate 3  & 2 & $1366.8027_{-0.0027}^{+0.0057}$ &$1366.8028_{-0.0023}^{+0.0034}$\\
Candidate 3  & 3 & $1373.3580_{-0.0070}^{+0.0055}$ &$1373.3614_{-0.0033}^{+0.0036}$ \\
Candidate 3  & 4 & $1379.9152_{-0.0048}^{+0.0042}$ &$1379.9154_{-0.0027}^{+0.0025}$\\
\hline 
\hline
\end{tabular}
\end{center}
\label{tab:posterior_ttvs}
\end{table}%

 \begin{figure}
\begin{center}
\includegraphics[width=1\linewidth]{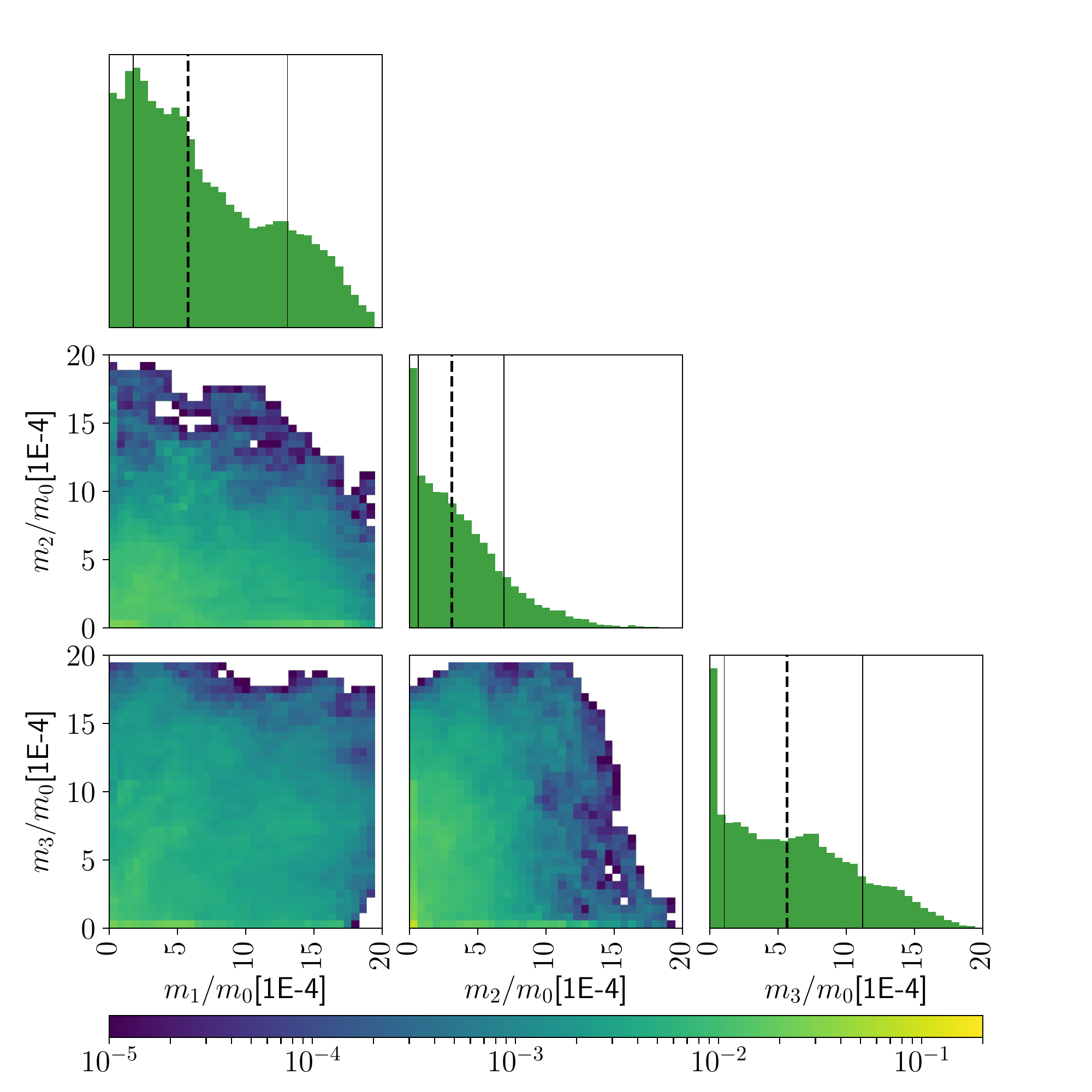}\\
\includegraphics[width=1\linewidth]{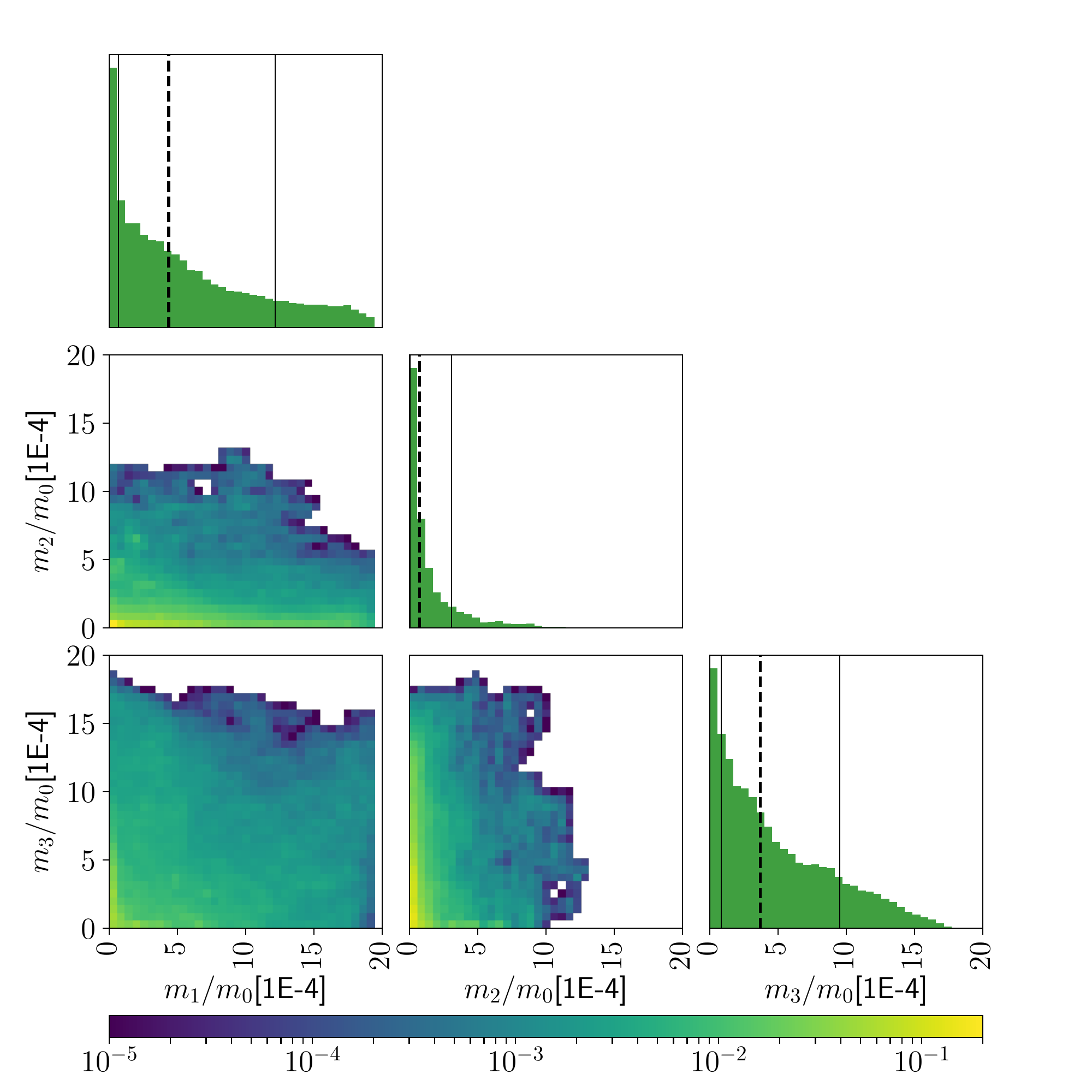}\\
\caption{\label{fig:MA1} Posterior distribution functions of the planetary masses for the scenario (i) on the top panel, and scenario (ii) on the bottom one. The histogram subplots on the diagonal show the median (dashed line) and the one-sigma uncertainty (thin lines) of the estimated masses.}
\end{center}
\end{figure}

\end{document}